\documentclass[aps,prl,twocolumn,superscriptaddress,groupedaddress]{revtex4}
\usepackage{subfig}
\usepackage{graphicx}  
\usepackage{dcolumn}   
\usepackage{bm}        
\usepackage{amssymb}   
\usepackage{slashed}
\usepackage{graphicx}				
\usepackage{amsmath}
\usepackage{mathtools}
\usepackage{tikz,pgf}
\usepackage{comment}
\usepackage{asymptote}
\usetikzlibrary{arrows,backgrounds}
\usetikzlibrary{fit,scopes,calc,matrix,positioning,decorations.pathmorphing}
\usepackage[all]{xy}
\usepackage{yfonts}

\newcommand{\ket}[1]{\ensuremath{\left|#1\right\rangle}}

\begin{document}
\title{Ancilla mediated higher entanglement as T-duality, a categorial conjecture}
\author{Andrei T. Patrascu}
\address{ELI-NP, Horia Hulubei National Institute for R\&D in Physics and Nuclear Engineering, 30 Reactorului St, Bucharest-Magurele, 077125, Romania}
\begin{abstract}
Using a higher categorial interpretation of entanglement involving gauge theories and $\sigma$-models instead of qubits, one recovers T-duality as a form of ancilla aided entanglement generation. This opens the way towards new dualities in gauge theories and $\sigma$-models produced by means of analogies with quantum circuits of various types. 
\end{abstract}
\maketitle
\section{Introduction}
Both special and general relativity are pillars of our modern understanding of reality. They were fundamentally classical theories, as quantum mechanics has only been introduced decades after their construction. Being constructed without the constraints of quantum consistency, they became also particularly difficult to reconcile with the quantum structures. Special relativity has been successfully reconciled with quantum mechanics via quantum field theory and the process of renormalisation. General relativity resisted a similar approach due to its non-renormalisability and its ill UV behaviour[1-4]. However, various dualities have shown that classical gravitational solutions encode precisely properties of quantum field theories in different regimes. One can therefore ask if there is some quantum "remnant" in general relativity that makes such descriptions possible. The ER=EPR proposal links two black holes connected via a wormhole to quantum entanglement [5]. The AdS/CFT duality links a classical anti-de-Sitter solution of general relativity to a highly quantum conformal field theory in a different regime. Lacking a proper axiomatisation of quantum mechanics, one has to rely on the most common quantum properties that differ from classical physics in order to derive what the quantum "remnants" could be that can be found in general relativity. There are several ways in which quantum mechanics differs from classical mechanics. First of all, observables are expanded to the status of operators acting on a Hilbert space of states. The outcome of measurements is not anymore one single possible result, but instead one has to consider all possible outcomes and to propagate them accordingly via a wavefunction or, further on, in quantum field theory, via a path integral quantisation prescription. As observables become operators, one can define a notion of compatibility between them, and hence a notion of commutativity. Incompatible observables will not commute, leading to what we understand as quantum fluctuations. The state of the system is described by means of a state of maximal knowledge [6], through a wavefunction, and not by means of a state of complete absolute knowledge, as was the case in classical mechanics. Perfect knowledge was assumed in classical mechanics even at the statistical level, where fluctuations were caused by subjective limitations and not by fundamental uncertainty. Finally, probably one of the main properties of quantum mechanics that I believe will resist most reductionist attempts implicit to axiomatisation, is the non-cartesianity of the spaces of states. While in classical physics, an object is constructed from individual separated pieces that are placed in a whole and allowed to interact in a rather newtonian way, and we can always separate the system back into its individual pieces without losing any information, this becomes impossible in quantum mechanics, where the global wavefunction of the whole system encodes more information than any individual wavefunction associated with any of the pieces. Moreover, the informational difference between the state space of the combined system and that of the individual pieces put together in a cartesian way grows exponentially, forcing us to use the tensorial pairing available in quantum mechanics. In essence, we have a tension between a local and a global interpretation of physics, local referring to the underlying subsystems, and global referring to the over-arching combined system. Moving between scales as done in the prescriptions of the renormalisation group does something similar. We integrate over small distance degrees of freedom producing effective theories depending only on larger distance degrees of freedom. This type of decoupling the small scale degrees of freedom does not completely eliminate the effects of the small degrees of freedom. In fact, their existence manifests itself in the requirement of seeing the couplings of the theory as variable, depending on the scale at which we perform the measurement. However, from this point of view, we do obtain a consistent effective theory if we insist to perform measurements at a given energy scale. More interesting things happen if we have relevant degrees of freedom (say a scalar particle like the Higgs boson) that are fundamentally dependent on the physics at or beyond the cut-off scale of the theory. Moreover, the renormalisation group may also have topological features leading to global structures connecting different scales, making an easy decoupling impossible. The fact that the apparent decoupling between scales that appears in usual quantum field theories occurs is in a sense still quite mysterious, although of course we can see it happening by performing the calculations. We do however have to remember that the decoupling between scales is never absolute, and in reality we can never really completely forget about the physics of the lower length (higher energy) scales unless we decide to ignore certain questions, particularly those related to naturalness, the hierarchy problem, the values of various particle (or resonance) masses, coupling constants, the cosmological constant problem, etc. 

\par From this discussion an important aspect emerges: the concept of maximal knowledge, encoded by the wavefunction as mentioned by Schrodinger in his original discussion on the state of quantum mechanics [11] is a very general idea that must take into account information in all its possible forms. For a long time, starting with the ideas of Turing and Shannon, information has been regarded in a practical sense, as certain "markings" made on a substrate. The markings were usually assumed to be objects with two possible states, like 0 and 1, and the substrate was always considered to be some physical structure, from the classical example of a tape, a magnetic tape, or a disk, to the mode elaborate states of quantum systems where superpositions between the two possible states were also allowed and the substrate became more fundamental, quantum states of atomic or nuclear spins, topological states, etc. 
In all these cases however, the assumption was the same: information is represented as "markings" on a "tape". However, category theory teaches us that often "maximal knowledge" (to use Schrodinger's terminology) can be extracted without direct reference to the object about which we wish to extract knowledge. In fact, category theory allows for a perfect characterisation of objects by means of so called "universal properties" that refer to their "surroundings" and not to the objects themselves. Much information can be assigned to objects by looking at their context instead of the information that could be encoded "inside" them. This makes sense in category theory, where objects themselves can be regarded as devoid of inner structure and yet carry significant properties. 
Another situation where this approach could be of major importance is the inner structure of field theories or of "particles". The existence of space-time symmetries associated to a certain system, in particular, a point-like particle, is well known since Einstein and is represented in the form of the so called Poincare group (in the case in which the spacetime is assumed flat and one can assume to a certain degree of accuracy the existence of translation invariance). The existence of "inner" symmetries however appeared as a necessity due to various discoveries in the early 20th century. For one, the existence of a semi-integer spin (fermions) required the assumption of an interesting inner structure which had to be associated to higher homotopy groups. In fact, the existence of spinors or a spin structure on a manifold depends on the triviality of the Stiefel Whitney class of the manifold. A non-trivial such class implies an obstruction to the existence of a spin structure. In effect a point like particle is assumed to behave under rotations in a way that retains the memory of the type of rotations that have been performed on it, in particular whether the paths followed in the rotation group can be continuously deformed into each other. Even if the end-result of two rotations is the same, this can be achieved through inequivalent combinations of rotations, resulting in an overall change of sign. Our point-like particle has a "memory" of those particular rotations performed on it and can ultimately tell whether one went on one of the paths or the other by its overall sign. Simply put, that means that the spin group is a double cover of the rotation group, as obviously each rotation can be produced in two inequivalent ways as the final outcome of a group path. While vector or tensor quantities cannot detect such double cover and hence the end position of a set of rotations represented by a path in the rotation group will not depend on the path itself, in the case of spinors, we have sensitivity to such rotation group paths, and hence to the homotopy class of the rotation path. This sensitivity implies the fact that our point-like fermions perceive the topological structure of the rotation group in the manifold where they exist. To be more precise, a point-like object detects global topological features associated to an extended object or at least an object that should have some form of "inner structure". 
This is of course not the only example where this "inner structure of a point" becomes significant. In fact, quite a bit of "inner symmetry" has been discovered in various theories, in particular in the standard model of elementary particles. 
Historically, a series of no-go theorems have shown how the inner symmetries of a point are impossible to be combined in a non-trivial way with the Poincare symmetry. This amounted to the development of the so called Coleman Mandula theorem which claimed that the only way in which the Poincare algebra can be combined with the other inner symmetries is by means of a direct product. The consideration was of course that the only symmetry algebras are Lie algebras. 
This theorem was famously superseded by the Haag-Lopuszanski-Sohnius theorem which was a generalisation of the Coleman-Mandula theorem to Lie superalgebras. It introduced Lie superalgebras as the only new spacetime dependent symmetry that allows for a non-trivial combination of spacetime symmetries and inner symmetries. This is a result of inclusion of commuting and anti-commuting generators, where the anti-commuting, fermion generators are key in allowing for non-trivial combinations of spacetime and inner symmetries. 
We noticed however, that the fermion structure (or spin structure) appears as a gained sensitivity to a higher homotopy group of the rotation group. If additional sensitivity to such global structures can be associated to point-like objects then further extensions of the ways in which inner symmetries and Poincare symmetries can be combined should emerge. 
String theory is based on the idea that an extended object would clearly have such properties. However, the same can be achieved by considering "external memory" or an "external record" of how the path in the rotation group was performed. This could be seen by means of the so called belt trick, and its interesting application is by linking our point-particle via one (or several) such "belts" to fixed external references. What is essential here is that the "memory" of the path in the rotation group doesn't have to be the object itself, but can also be represented in all possible ways in which it can be connected to outside regions (say, an external memory). Designing a categorical construction, it would mean that by certain universal structures, properties of objects can be inferred (and attached to the objects) that would remain point-like and would not be able to determine any inner structure. 
A metaphorical, but simple example would be the information about the content of a book. That information could be retrieved by reading the book, but a significant amount of information could be obtained by looking at the section of the library where the book has been placed, the year of its writing, the style of the author, the title, the cover, the literary style, the predominant theme of the period, etc. All in all, much information can be obtained without even reading the book (following symbols on a tape) if the book is properly placed in its context. 

\par Another aspect that is common to quantum computation but not so much to classical computation is the concept of reversibility. Quantum processes need to be reversible. In classical mechanics a bit of memory can be turned on or off (flipped) without considering its history. This is impossible in quantum mechanics as the qubit operations need to be reversible. We find this in another situation in classical physics, namely in the construction of spacetime worldsheets or worldlines and in the cobordism category that defines their interaction [7]. This is why a potential functorial relation between the category of worldsheet cobordisms and the category of Hilbert spaces with their operations has been postulated. In quantum information we will need ancilla qubits to encode certain states with desired properties. This is necessary because in general we cannot act on a qubit while ignoring its history or "un-doing" actions on qubits. This property will become important in what follows. There are at least three domains where the global vs. local tension becomes interesting. First, it is renormalisation. The idea of renormalisation is based on the decoupling of scales. Basically, the attempt at renormalisation is an attempt at systematically decoupling the different scales of energy such that one can construct a meaningful theory at a certain scale, while incorporating the other scales in some effective manner. The effective contribution of the smaller scales usually manifests itself as flow equations in the parameters of the effective theory. That is also why we need to apply certain operations to the couplings of our theory as we move from one scale to the next (the well known renormalisation (semi)-group equations). The redefinitions of parameters and terms in the case of the standard model interactions are finite and preserve the predictive power of the theories. They do represent an elegant way of connecting different scales, and of combining subsystems and constituents into a larger global system. The problem of describing strongly entangled/coupled systems in this way is not fully resolved in more than one dimension [8]. The origin of this problem resides in the exponential complexity of the many body wave function. I discussed previously the various results emerging from a topological analysis of the renormalisation group and the problem of whether a fibre bundle approach to the RG transformation could be a suitable approach is still open [9]. In any case, the main problem originates from the fact that indeed the transition between scales is not as simple as one may think. While we can renormalise all interactions of the standard model, adding gravity to the system makes the renormalisation procedure ill defined. Indeed, as is well known, the renormalisation group is not actually a group, it is more like a semi-group, as its inverse is not defined. Indeed, in that sense if one takes into account topological features of the renormalisation one is well advised to consider principal bundles where a preferred choice of identity is non-existent. The topological non-triviality of such a scale transformation would imply the non-existence of a trivial cartesian product between domains of scale, and hence would lead to scale interference, which is expected in string theory due to the T-duality and in the standard model via the flow of couplings [10]. Of course the two effects are apparently rather different. In any case, they are unified by the idea that the global structure cannot directly be un-paired into constituent pieces. 
The other domain where the local/global tension emerges is of course quantum mechanics itself, where we have entanglement which is a property of a combined super-system, which has information on the subsystems that cannot be recovered when analysing the subsystems separately. 
The third domain where we have a local/global problem is gauge theory. There we try to develop theories with consistent interactions, only to find that the interactions emerge from a fibre bundle approach based on a structure group which we associate to a gauge symmetry, and that this bundle encodes significant global information. In fact, a measure of curvature is what makes interactions possible, hence a global property, which cannot be detected at each point in space separately lies at the foundation of gauge interactions. Indeed, one of the first principles of general relativity was that there always exists a transformation that makes the spacetime locally flat, hence a freely falling observer, unless having access to some global information would not know it is in free fall. In order to obtain a theory of gravity, one had to incorporate this global information by means of curvature and that had to be defined by means of a connection on the fibre bundle.
Indeed curvature and topology cannot be fundamentally separated. Curvature influences topology and reversely. 
 It comes at no surprise that quantum wavefunctions and quantum fields also have fibre-bundle interpretations, however they are less often used due to us being accustomed with other tools through which we link local and global structure, for one, the partition function and its path integral approach to quantisation, and the combination of phases of wavefunctions. Wavefunctions interpreted in a fibre bundle context rely on the prescription of geometric quantisation. Given the phase space and a symplectic form, the line bundle over the phase space equipped with a U(1) connection is defined such that the curvature of this connection is the symplectic form. A choice of polarisation over the phase space produces a split between the coordinates and momenta, leaving the quantum wavefunctions to be the sections of this line bundle that depend only on the position, therefore are constant on surfaces of constant coordinate. The evolution is given by the unitary Hamiltonian encoded in the symplectic structure acting on the space of states. However, if we admit that an essential property of quantum mechanics is that it takes into account non-cartesianity and hence that separation of a quantum system in local subsystems is often incomplete due to entanglement, we have to also admit that both the other problems also have a quantum nature. We therefore have an additional bundle structure that encodes the entanglement properties, and this is the subject of our discussion here. Indeed, not any wavefunction is entangled. Entanglement emerges from the non-separability of the constituent states after the larger system is formed from the separate subsystems. After the quantisation has taken place, or, if we have already a quantum theory, it is this non-cartesianity amounting to the impossibility of simply describing a composed quantum system in terms of cartesian pairs of its subjacent structure, that plays the important role in defining the bundle structure I will consider next. 
\par While nobody questions the idea that renormalisability is a feature of quantum field theory, the fact that general relativity incorporates a similar tension between local and global structures as the one found in quantum mechanics did not seem to have occurred until now. If this is accepted, then general relativity already has a property of quantum mechanics, namely non-cartesianity, incorporated in its construction: one cannot infer the same information from local measurements as one can from global ones. Also, one cannot operate on a qubit (say flip it) while neglecting its history, in the same way in which one cannot act upon a world-line event disregarding its history. It simply so happens that the mathematics chosen to describe this tension was based on fibre bundles with connections, and this geometric approach is used more often in the general relativity context, albeit trying to solve the same problem as the partition function and the path integral quantisation tries to solve in quantum field theory. 
\par This property is of course not the sole property an axiomatisation of quantum mechanics would contain. Indeed, in quantum mechanics we have to expand our observables from objects that can take one single possible outcome and hence imply absolute knowledge, to operator-valued observables which rely on several possible outcomes having various probabilities of emerging at different times when the experiment is repeated. But this is also a method through which one gains access to the global structure of the state space, where different potential outcomes combine to generate a statistics that takes such potential events into account. Hence, even this construction was required because one had to expand from a local viewpoint of outcomes to one that involved several successive repetitions that contribute and interfere to produce the overall outcome. Such observables may not commute, meaning the information they encode globally is not locally compatible. But a very similar aspect occurs in general relativity. In fact the connection is used to compare space-time separated objects in a consistent way, while the curvature of a space being defined by the non-commuting property of covariant derivatives resulting from the existence of a non-trivial connection. We use the non-commuting property of connections (or covariant derivatives) to get a measure for some global property, namely curvature. This is surely not the sole global property one can consider, but if one considers only point particles, one is less sensitive to more interesting non-local properties of spacetime. There should be analogues of such non-local properties also in quantum mechanics. So, here we have, yet another property that general relativity shares with quantum mechanics, and it also originates from the local/global tension that appears in both domains. One attempt would be to construct two categories, one of quantum physics, and the other of general relativity, and to notice how both could be related and what functors one could develop to relate them. Indeed, this would show that the two are in many ways similar. One could therefore say that general relativity is already quantum to some extent, or that it preserves some quantum remnants from its overarching quantum sibling, potentially string theory. One may be surprised therefore at the difficulty of completing the UV domain of quantum gravity. After all, other theories that share similar properties have a well defined UV limit, say quantum chromodynamics (where of course the problems appear in the low energy limit where the couplings are strong, but that is another subject not to be discussed here, however think at asymptotic freedom and asymptotic safety). The solution to the UV problem of quantum gravity comes from incorporating extended objects instead of points. Of course extended objects would make the connection between local and global properties more amenable to a construction that is based on quantum mechanics. Once this is realised one can continue with the tools and methods of string theory and describe quantum gravity. However, various limits in string theory give us links between so called classical general relativity solutions and non-stringy quantum field theories. This can only be possible if general relativity retains some of the quantum properties of string theory. 
\par An important result of string theory and compactification is the fact that the couplings of a specific quantum field theory may have a flow that is determined by the equation of motion of the moduli of a string compactification. In this sense, we may discuss sigma models in which the couplings of the theory are represented as tensorial quantities (or spinorial quantities for that matter) that exist in a so called target space $\mathcal{M}$. in what follows I will cover some very well known material related to $\sigma$-models. I will follow the path of established understanding and I considered one good reference for this path to be [11]. There are of course several other books and articles on this topic, as the subject is, as said, well established. I however used ref. [11] and I found it very suitable to the tasks of this paper, hence I followed it in considerable detail for this introduction. The target space of a $\sigma$-model has the properties of a topological and metric space and the coefficients represent transformations of the fields of our theory. In there, the fields of the theory represent coordinates, while the couplings represent transformations of such coordinates with different properties. In this space, if string models are accepted and compactifications are performed, the parameters describing such geometric and topological structures are in fact determining the dynamics of such tensor couplings. Therefore, the equations of motion of the geometric parameters of the target space are to be associated to flow equations of the couplings. It was therefore assumed in the compactification program that such geometric properties will ultimately determine the parameters of the initial quantum field theory. However, this unfortunately didn't come to be, not because this approach was wrong, in fact the idea of providing a topological and metric representation of the couplings may help understanding many of their properties, but because physics seems to have some properties even at the level of the "target space" that are not known. 
It is of course not only the target space itself that plays a role. The target space $\mathcal{M}$ has as coordinates only the scalar fields of our theory. However, on this space, various bundle structures exist. In fact, all non-scalar fields live in a bundle on top of this target space
\begin{equation}
\mathcal{V}\rightarrow \mathcal{M}
\end{equation}
If we consider only a scalar field theory with $D$ spacetime dimensions and denote the scalar fields $\phi^{i}$, $i=1,2,...,n$, and we write a local, Hermitian, Poincare invariant Lagrangian with at most two derivative terms, then we obtain 
\begin{equation}
\mathcal{L}=-\frac{1}{2}g_{ij}(\phi)\partial_{\mu}\phi^{i}\partial^{\mu}\phi^{j}+...
\end{equation}
where the dots indicate terms with no derivative. 
The greek indices go over spacetime and the latin indices go over the "coordinates" in the field space (namely our scalar fields). 
The coupling of derivative terms is a real symmetric field dependent matrix $g_{ij}(\phi)$. Due to unitarity the kinetic terms must be positive and hence our matrix must be positive definite. Physical quantities should not depend on field reparametrizations 
\begin{equation}
\phi^{i}\rightarrow \tilde{\phi}^{i}(\phi)
\end{equation}
and the Lagrangian becomes 
\begin{equation}
\mathcal{L}=-\frac{1}{2}\tilde{g}_{ij}(\tilde{\phi})\partial_{\mu}\tilde{\phi}^{i}\partial^{\mu}\tilde{\phi}^{j}+...
\end{equation}
with
\begin{equation}
\tilde{g}_{ij}(\phi)=\frac{\partial \phi^{k}}{\partial \tilde{\phi}^{i}}g_{kl}\frac{\partial{\phi^{l}}}{\partial\tilde{\phi}^{j}}
\end{equation}
We can therefore regard $g_{ij}$ as a Riemannian metric on the field space (manifold) $\mathcal{M}$. 
A classical field configuration would then appear as 
\begin{equation}
\Phi:\Sigma\rightarrow \mathcal{M}
\end{equation}
where $\Sigma$ is the spacetime which we may consider to be Minkowski (but that is not necessary) and the target space $\mathcal{M}$ which is a more complicated space that could even have non-trivial topology, and on which the scalar fields are coordinates. 
The Lagrangian then appears as the trace with respect to the spacetime metric of the pullback of the metric $\Phi^{*}g$. 
If we want to construct the renormalisation group beta functions in a theory with action defined as 
\begin{equation}
S=-\frac{1}{2\hbar}\int_{\Sigma}\sum_{l=2}^{\infty}g_{i_{1}i_{2}...i_{l}}\phi^{i_{3}}\phi^{i^{4}}...\phi^{i_{l}}\partial_{\mu}\phi^{i_{1}}\partial^{\mu}\phi^{i_{2}}d^{2}z
\end{equation}
(considering the Taylor coefficients of the metric $g_{ij}(\phi)$) we can construct an infinite set of beta functions which we can assemble in a symmetric tensor $\beta_{ij}(\phi)$ associated to the couplings $g_{i_{1}i_{2}...}$. The RG equations are therefore 
\begin{equation}
\mu\frac{\partial}{\partial \mu}\frac{g_{ij}(\phi)}{\hbar}=\beta_{ij}(\phi)
\end{equation}
The beta function is expected to be made up of the metric on the field space and its derivatives. However, we know that the beta function vanishes when the QFT is free, and hence when the metric is flat. Because of this observation we can write the beta function as an expansion in terms of the Riemann tensor and its covariant derivatives. 
Considering also some scaling properties with respect to the volume we obtain 
\begin{equation}
\beta_{ij}=c_{1}R_{ij}+c_{2}g_{ij}R
\end{equation}
In the case of compactification, the geometry and the topology of the field space will become highly nontrivial and the equations governing the variations of the parameters describing such compactifications will enter in the equation as dynamical objects with equations of motion. 
This definition is purely geometrical and of course involves also topological features. However, there are additional properties that can be associated to such dynamical objects. In fact, supersymmetry implies that the couplings are not independent as they would appear in a Lagrangian that contains scalar and fermion degrees of freedom. Supersymmetry would make certain selections of the specific bundles (associated to fermions) that can be constructed over the field space. In supersymmetric theories SUSY relations between different couplings are to be associated to relations between different geometric structures on our manifold $\mathcal{M}$. In all theories, the associated couplings are related to various geometric structures and objects that can be associated to or constructed over our manifold $\mathcal{M}$. What supersymmetry does is to define certain main geometrical structures that can exist on $\mathcal{M}$ and introduce new connections between those structures. 
It is probably important to describe briefly in this introduction how such structures come to be. This is so because the main goal of this article is to show that such geometric structures can be naturally related in ways that do not imply supersymmetry but in fact imply a higher form of entanglement (and/or correlation) between such geometric structures and their dynamical parametrizations. An entanglement at this level, which I call "higher entanglement" clearly produces a series of new dualities and relations between geometric structures that were not visible previously. In fact, the introduction of entanglement is a physical requirement, as nature seems to behave quantum mechanical in most contexts, and hence would not be visible from a purely geometrical and/or topological view of the manifold $\mathcal{M}$. 
\par We return therefore to our RG flow equation and analyse them more carefully. The fixed points of the RG flow are not necessarily zeros of the beta function, and in fact can correspond to a flow that acts on the metric as a diffeomorphism which would imply that the action is scale independent up to field redefinitions. In order to achieve this we write 
\begin{equation}
\beta_{ij}=\mathcal{L}g_{ij}
\end{equation}
where $\mathcal{L}_{v}$ denotes a Lie derivative along the vector field $v$ on $\mathcal{M}$. If the invariance of our Lagrangian field theory is represented as a continuous symmetry group $G$ acting on the field space, it should leave the metric invariant. Therefore $G$ should be a subgroup of the isometry group of our field manifold $Iso(\mathcal{M})$. 
We can introduce corresponding infinitesimal symmetries 
\begin{equation}
\phi^{i}\rightarrow \phi^{i}+\epsilon^{A}K_{A}^{i}(\phi)
\end{equation}
which are generated by the vector fields $K_{A}^{i}\partial_{i}$ with $A=1,2,...,dim(G)$ satisfying the Killing condition 
\begin{equation}
\mathcal{L}_{K_{A}}g_{ij}=\nabla_{i}K_{Aj}+\nabla_{j}K_{Ai}=0
\end{equation}
and the algebra
\begin{equation}
\mathcal{L}_{K_{A}}K_{B}=[K_{A},K_{B}]=f_{AB}^{C}K_{C}
\end{equation}
where we use the standard terminology and call $f_{AB}^{C}$ the structure constants of the Lie algebra of the group $G$. 
The Killing vectors $K_{A}^{i}\partial_{i}$, defining the geometry of $\mathcal{M}$, prescribe also the minimal coupling of the gauge vector fields $A^{A}_{\mu}$ to the scalars. To gauge the symmetry produced by a subgroup $G$ of the isometry group, we replace in the Lie derivative $\mathcal{L}$ the ordinary derivative with the covariant one
\begin{equation}
\partial_{\mu}\phi^{i}\rightarrow D_{\mu}\phi^{i}=\partial_{\mu}\phi^{i}-A_{\mu}^{A}K_{A}^{i}
\end{equation}
then we obtain infinitesimal gauge transformations 
\begin{equation}
\begin{array}{c}
\delta \phi^{i}=\Lambda^{A} K_{A}^{i}\\
\\
\delta A_{\mu}^{A}=\partial_{\mu}\Lambda^{A}+f^{A}_{BC}A_{\mu}^{B}\Lambda^{C}\\
\end{array}
\end{equation}
The parameters $\Lambda^{A}$ are arbitrary spacetime functions. 
We obtain as a consequence the transformation rule of the covariant derivative as
\begin{equation}
\begin{array}{c}
\delta D_{\mu}\phi^{i}=\Lambda^{A}(\partial_{j}K_{A}^{i})D_{\mu}\phi^{j}+\\
\\
+A_{\mu}^{B}\Lambda^{C}[K_{B}^{j}\partial_{j}K_{C}^{i}-K_{C}^{j}\partial_{j}K_{B}^{i}]-\\
\\
f_{BC}^{A}A_{\mu}^{B}\Lambda^{C}K_{A}^{i}=\\
\\
\Lambda^{A}(\partial_{j}K_{A}^{i})D_{\mu}\phi^{j}\\
\end{array}
\end{equation}
The closure of the gauge algebra implies the covariance of the covariant derivative. At the same time, the invariance of the kinetic term $g_{ij}D_{\mu}\phi^{i}D^{\mu}\phi^{j}$ demands $\mathcal{L}_{(\Lambda^{A}K_{A})}g_{ij}=0$, and we obtain 
\begin{equation}
\delta(g_{ij}D_{\mu}\phi^{i}D^{\mu}\phi^{j})=\Lambda^{A}(\nabla_{i}K_{Aj}+\nabla_{j}K_{Ai})D_{\mu}\phi^{i}D^{\mu}\phi^{j}
\end{equation}
We can see that the physics of a gauge $\sigma$-model is given to a large extent by the geometry of the target manifold $\mathcal{M}$. The target space seems to have similar properties with the spacetime manifold, namely it recovers the property of general reparametrisation invariance and we can therefore identify a target space equivalence principle. Any physical quantity, local in $\mathcal{M}$ and depending only on the metric and its first derivative is computed using a flat target space. 
The couplings of the scalar fields to gauge vectors are determined by a set of vector fields $K_{A}$ on the target manifold satisfying differential geometric constraints. In fact, all couplings in the Lagrangian can be associated to differential geometric structures on the target manifold. 
If we consider a theory with scalars $\phi^{i}$ and fermions $\psi^{a}$, with the indices $i=1,2,...,n$ and $a=1,2,...,m$ and we consider a two dimensional case (say worldsheet theories) we can construct a general Lagrangian of the form 
\begin{equation}
\begin{array}{c}
\mathcal{L}=-\frac{1}{2}g_{ij}(\phi)\partial_{mu}\phi^{i}\partial^{\mu}\phi^{j}+b_{ij}(\phi)\epsilon^{\mu\nu}\partial_{\mu}\phi^{i}\partial_{\nu}\phi^{j}+V(\phi)+\\
\\
+i h_{ab}(\phi)\bar{\psi}^{a}\gamma^{\mu}\partial_{\mu}\psi^{b}+ i\tilde{h}_{ab}(\phi)\bar{\psi}^{a}\gamma_{3}\gamma^{\mu}\partial_{mu}\psi^{b}+\\
\\
+k_{abi}(\phi)\bar{\psi}^{a}\gamma^{\mu}\psi^{b}\partial_{\mu}\phi^{i}+\tilde{k}_{abi}(\phi)\bar{\psi}^{a}\gamma^{\mu}\gamma_{3}\psi^{b}\partial_{\mu}\phi^{i}+\\
\\
+y_{ab}(\phi)\bar{\psi}^{a}\psi^{b}+\tilde{y}_{ab}(\phi)\bar{\psi}^{a}\gamma_{3}\psi^{b}+\\
\\
s_{abcd}(\phi)\bar{\psi}^{a}\psi^{c}\bar{\psi}^{b}\psi^{d}+...\\
\end{array}
\end{equation}
with each coupling being a function of the scalar fields (coordinates) $\phi^{i}$. The important aspect of this representation is that every term and hence every coupling in this Lagrangian is seen as a geometric structure on the target manifold. The function $g_{ij}(\phi)$ is a Riemannian metric, the coupling $b_{ij}(\phi)$ is a differential 2-form $b=\frac{1}{2}b_{ij}(\phi)d\phi^{i}\wedge d\phi^{j}$, $V(\phi)$ is just a scalar field on $\mathcal{M}$. If we go to the Majorana-Weyl representation, the chiral fermions $\psi_{\pm}^{a}$ are sections of vector bundles over the spacetime $\Sigma$ given a field configuration $\Phi:\Sigma\rightarrow \mathcal{M}$. They are pull-backs $\Phi^{*}\mathcal{V}_{\pm}$ of real vector bundles $\mathcal{V}_{\pm}\rightarrow\mathcal{M}$ with fibre metrics $h^{\pm}_{ab}$, etc. 
In general, fields with non-zero spins can be seen as sections of vector bundles of the form 
\begin{equation}
S_{R}\otimes \Phi^{*}\mathcal{V}\rightarrow \Sigma
\end{equation}
where $S_{R}\rightarrow \Sigma$ is regarded as a vector bundle associated to the spin representation $R$ of the respective field, and $\mathcal{V}$ is a target space bundle that characterises the interaction of that field with the scalars. The derivative couplings are determined by the covariant derivatives with associated connection being the natural connection on the bundle, etc. 
We therefore see that in all theories, the couplings are described by geometric objects and that such objects are a-priori independent of each other. 
Supersymmetry acts as a selector of the specific geometric structures that are compatible with it, and moreover, due to supersymmetry, new connections between such geometric structures are implies. 
However, and this is the main subject of this article, there exist other types of connections, not initially represented in a form of symmetry, but rather in the form of a correlation between such geometric structures, and in particular, if a wavefunction description is understood on this target space, there can exist in fact a higher form of entanglement that correlates geometric structures. The existence of such a new higher form of quantum correlations is what this article constructs by identifying first, potential entanglement between auxiliary or ghost fields. As it is clear that such fields would not appear as exterior fields associated to some measurable particles, I am not however concluding that because of that, there is no quantum structure and higher correlation to be found among those fields. In fact, there should be, and such entanglement should be associated to some dualities in ways that were not known previously. 

\section{Auxiliary fields}
Auxiliary fields are essential in formulating a series of theories, from supersymmetric theories, to gauge theories etc. in the context of quantum mechanics. The reason why these auxiliary fields are important is because they are literally used to keep track of certain properties that do not appear in instances where the theory is naively realised. To be more specific, we need to introduce auxiliary fields in order to keep track of the chiral components of on-shell Weyl fermions. Why do we need that? Simply put, off shell, the Weyl fermions are not really Weyl fermions anymore. They gain additional fermionic degrees of freedom and are described by a Dirac fermion with four degrees of freedom, not by the two degrees of freedom of the Weyl fermions. Because of this, if we tried to describe a new symmetry (say, accidentally supersymmetry) on shell, we will notice that its algebra will not close. Therefore, it is required to modify our initial symmetry transformations to account for the additional off shell degrees of freedom by introducing auxiliary fields that vanish by the equation of motion on shell. The same situation happens for gauge degrees of freedom and in other situations, for example in the case of BRST quantisation. A more general situation is the Batalin-Vilkoviski method of quantisation or the field-anti-field approach. Given that such fields are essentially non-dynamical (from the perspective of on shell physics) and they do in fact have no effect once they are eliminated by integration (and by specifying the equation of motion on-shell, depending on the case), we often call this prescription as "a trick". 
Such a "trick" however is not as inoffensive as one may think. It works fine, by all means, but it is an indicator for additional global effects that happen off-shell and do in fact impact our on-shell analysis, even if it just amounts to transformations of the presupposed algebras by means of the auxiliary fields (I am thinking at the F-fields in supersymmetry for example). 
What happens is that additional degrees of freedom, vectorial, fermionic, etc. happen to have an impact on how we have to define our symmetries (if we want to define them) although the fields themselves are irrelevant on-shell. 
In this article I take those auxiliary fields more seriously and ask whether they have additional quantum properties that the simple Lagrangian formulation of our theories does not immediately take into account. In fact, I consider that such auxiliary fields may in fact not only have quantum properties, but be even entangled and this entanglement, while not obviously manifest or visible off shell, produces effects that are non-local in nature, couple several or any scales and are extremely important in our theories, but are lost simply because our theories are relying on simple or semi-simple compact groups and their unique associated Lie algebras. Entanglement effects of degrees of freedom that appear off-shell only amount to modifications of the associated group structure, making it potentially not even a group, but a formal group, with totally different and non-trivial global structure. 
Beyond this approach, the off-shell structure may involve higher gauge symmetry and the associated categorification may imply the existence of higher field forms coming together with their entanglement. 
The main idea is to present side to side two constructions, one in which we create an entangled system, and we bring that discussion up the point of obtaining an entangler gate by means of a series of ancillas and the other one is one in which we obtain T-duality by means of the use of auxiliary fields and the integration over them. The parallelism between those two constructions leads us to the conclusion that something similar to entanglement is expected to occur at the level of the auxiliary (off-shell realised) degrees of freedom. 
Given such off-shell quantum properties, it is interesting to consider whether the off-shell region doesn't act in a more than trivial way on the constructions that can be realised on-shell, in particular by means of global effects and non-perturbative phenomena originating in quantum information theory. 

\section{T-duality as a result of a higher-entangled state}
We expect to have a functorial relation between general relativity and quantum mechanics, although it is not fully understood what exactly the functor should be. We do understand, following J. Baez and his categorial approach to the two theories, that the non-cartesianity of the two categories, that of the Hilbert spaces and the linear operators between them, and that of smooth manifolds and cobordisms are linked by some functor that allows us to interpret operations in one theory by means of analogue operations in the other. This approach leads us to the understanding that there is a deeper connection between the two, that we didn't fully explore until now. However, what seems to be particularly intriguing is how the two theories deal with global structure and particularly how they transition from local to global structures. While it is true that both do this and hence their structures are linked precisely by this, the mathematical tools used to link local and global structures were so different that a unification is still lacking. String theory, surprisingly has one component that makes this connection apparent, namely T-duality, which in some flux compactifications indeed links two string theories with different topology, and particularly, also links scales, transforming the trajectory of scale transformations from a straight line to a circle. In fact string theory solves many of the conundrums of quantum gravity, so much so that it has convinced many (and certainly this author) that if it is not the final theory of quantum gravity, it certainly walks the right path. In general T-duality links a theory with big spacetime radius with another one with small radius. In the case of one single compactified dimension with radius $R$ all physical properties of an interacting theory remain unchanged if one replaces 
\begin{equation}
R\rightarrow \frac{\alpha}{R}
\end{equation}
Of course this works if the dilaton field is transformed by 
\begin{equation}
\phi\rightarrow \phi- log(R/\sqrt{\alpha '})
\end{equation}
This can be expanded to the case of toroidal compactifications for a constant metric and an antisymmetric tensor as
\begin{equation}
\begin{array}{c}
(g+b)\rightarrow(g+b)^{-1}\\
\phi \rightarrow \phi - \frac{1}{2} log(det(g+b))
\end{array}
\end{equation}
The $T$-duality is generally regarded as somewhat mysterious. However, considering the connection between the fibre-bundle construction of gauge theories and of gauge connection on one side and the quantum mechanical entanglement as well as the emergence of spacetime (and in general geometry) from entanglement, the idea that the small and large scales of a theory with interactions should generate equivalent theories is not as surprising anymore. If indeed we have a functorial connection between quantum mechanics and gauge theory, as I expect to be the case, T-duality linking vastly different metric scales is just the reflection of the relation between long and short length entanglement. Being fundamentally a correlation and taking into account the global structure of the base manifold, entanglement should have the same properties, linking vastly different scales, hence T-duality should be nothing but a reflection of this nature of entanglement for sigma-models. Moreover, the causal structure, and hence the concept of spacetime interval is altered by a T-dual approach. 
Focusing on the nature of the pairing in quantum mechanics and gauge theory and on the functorial connection between the two, it is interesting to try to analyse T-duality from a quantum mechanics point of view, allowing for its quantum nature to emerge. Let us focus on the quantum result that the state space of a composed system is not reducible to the cartesian product of the state spaces of its subsystems. That implies that there exists a global structure to be accounted for. The non-trivial fibre bundles describe precisely how such a local cartesian product fails to be replicated globally. Fascinatingly enough when deriving the T-duality one obtains an exact analogue of quantum entanglement for gauge theories (or sigma models). Let us consider a manifold $M$ with metric $g_{ij}$ with $i,j=0,...,d-1$, the usual antisymmetric tensor $b_{ij}$ and the dilaton field $\phi(x_{i})$. The $\sigma$-model associated to it has at least one symmetry associate to an (in this case abelian) isometry on the metric. The procedure of obtaining the T-duality is to first gauge the isometry group, introducing the gauge field variables $A$. The field is required to be flat by means of a Lagrange multiplier $\xi\cdot dA$. We then perform two equivalent operations differing simply through the order in which they are performed. First we integrate over $\xi$, obtain a $\delta$ function $dA$ on the measure and the result of our constraint, namely that $A=d\psi$ is a pure gauge is being implemented. If we fix $\psi=0$ we recover the original model. Next we first integrate the gauge field $A$. We don't have a gauge kinetic term, hence we simply perform a Gaussian integration and obtain a Lagrangian depending on the original variables and the auxiliary variable $\xi$. We continue by fixing the gauge and obtain the dual action. It is important to highlight the aspects that are based on a quantum information understanding of the procedure leading to this duality. The novelty here lies in the interpretation, namely that T-duality can be seen as an effect of quantum entanglement in a categorial sense, applied to higher structures, leading to "entangled theories". This interpretation may shed new light on the origins of dualities in gauge and string theories. The first step is to lift the field space to a higher space, incorporating the gauge fields associated to the isometries of the metric. Then we require that for the resulting field, the gauge connection is flat, this condition being implemented by an additional field introduced in the Lagrange constrained as an auxiliary field, or an extended form of an ancilla quibit, encoding a technique of fixing part of the global structure. We take the inspiration from control ancilla-mediated quantum computation. In fact the method generating T-duality (at least for the abelian case) is analogous to the manipulation of registers via ancilla qubits leading to an entangled system. Basically in the first part we implement the construction of an ancilla entangled with the first set of fields in our initial theory. Another ancilla comes on the other side of the theory in the form of our condition of having a pure gauge field and a flat connection, implemented via another auxiliary field. Now, we integrate / measure the ancillas erasing the information regarding what theory is "entangled" to what ancilla. We obtain a fully entangled state between four objects. If next, we implement a "measurement" on each ancilla separately, we obtain an entangled state between the two regions. In quantum information the protocol is relatively simple. We construct the ancillas as
\begin{equation}
\ket{C_{\alpha}^{\pm}}=\mathcal{N}(\ket{\alpha}\pm \ket{-\alpha})
\end{equation}
with $\mathcal{N}_{\pm}=\frac{1}{\sqrt{2(1\pm e^{-2 |\alpha|^{2}})}}$. The two qubit measurement process of the two ancillas generated as a two-qubit measurement produces the global state, while the single qubit ancilla measurements make the entanglement manifest. What we obtain is an entangled state between the two sides with unit probability. 
The procedure is as follows: first we generate local entanglement between two non-interacting qubits, using ancilla qubits. The four qubits are initialised for example in the $\ket{+}$ state. Next a CPHASE gate is applied to the first qubit and its ancilla, followed by a Hadamard rotation of the first qubit. The same is applied on the second qubit and its ancilla. This generates an entangled state of the first qubit with its ancilla and the second qubit and its ancilla 
\begin{equation}
\begin{array}{c}
\ket{e_{1}}=\frac{1}{\sqrt{2}}(\ket{f,f}+\ket{a_{1},a_{1}})\\
\\
\ket{e_{1}}=\frac{1}{\sqrt{2}}(\ket{s,s}+\ket{a_{2},a_{2}})\\
\end{array}
\end{equation}
We then perform a two qubit measurement on the two ancillas. The result is a four qubit entangled state that, according to the outcome of the measurement is either 
\begin{equation}
\frac{1}{\sqrt{2}}(\ket{f,f,s,s} + \ket{a_{1},a_{1},a_{2},a_{2}})
\end{equation}
or 
\begin{equation}
\frac{1}{\sqrt{2}}(\ket{f,a_{1},s,a_{2}} + \ket{a_{1},f,a_{2},s})
\end{equation}
Performing single qubit measurements on the ancillas, the first and the second qubits become projected onto a particular entangled state, for example
\begin{equation}
\ket{\Psi}=\frac{1}{\sqrt{2}}(\ket{+,+}+p\ket{-,-})
\end{equation}
This has led to an entangled state between the first and the second qubit. 

While this procedure is not fundamentally new, it is important to understand how it is realised in a categorial interpretation of entanglement looking at gauge theories instead of quantum qubits. 

First, gauge theories have an entanglement remnant from their construction as fibre bundles that implement, generally non-trivial, topological structures that are not describable globally by means of cartesian products of local patches. 

Indeed, if one considers the procedure of generating a gauge field, hence "gauging a symmetry" one basically defines a gauge connection on a fibre bundle, and in order to do that, one requires the gauge group, which defines the global structure. This makes the patches that one wishes to connect on the base manifold, fundamentally impossible to decouple and decompose in separated, disjoint objects that would retain the global information. This is where the tension between local and global structures emerges and why we share the fibre bundle construction both in the case of gauge theories and in the case of a geometrical interpretation of quantum wavefunctions. The fact that our basis manifold in our derivation of T-duality has a gauge symmetry that we wish to gauge, introduces a global structure which we can use to create "entangled" theories, in a categorial sense. In this case however, we introduce a gauge field that is associated to a flat connection, and in order to do that we introduce also another field that implements this constraint. These are our ancilla constructions. We perform then two operations in different orders. What those realise is to eliminate the information between the two regions of the "entangled" theory that would be associated to each theory, and hence "entangles" the two sides of our construction. Then we Integrate over the respective ancillas individually, obtaining two theories that are dual. The two dual parts of the theory are represented by the two higher-entangled theories, hence it comes as no surprise that the resolution of one provides insight on the solutions of the other. 
Let us rephrase: We have two qubits that are not entangled. We bring in two ancillas, they represent two auxiliary fields in our construction, both having the role of gauging a symmetry of the worldsheet and implementing a constraint on the global structure (imposing a flat connection). By the way they are introduced they generate global information shared by the two sides of our to-be duality. Our flat connection condition generates the globally shared information. Now we can decide to integrate out and eliminate the two ancillas, each at a time, leading to two theories, which continue to carry the global information shared by the previous conditions implemented by the ancillas but now being directly entangled. This results in two theories that are said to be T-dual. This therefore shows that, in this interpretation of the gauging procedure, the resulting duality is basically a higher form of entanglement. 

It is not trivial to construct this type of higher entanglement. We need to give a pertinent definition of the operations we find in designing quantum information circuits that play a role in the operations we can perform on gauge theories and $\sigma$-models. This amounts to a consistent part of the argument being constructed by means of, what mathematicians would call, abstract nonsense, and as is usually the case with abstract nonsense, the results are deep and far reaching. In general quantum algorithms ancilla qubits are required because they need to maintain some level of reversibility of the computation. Indeed, quantum computation must be reversible, something that it has in common with the category of cobordisms in the worldsheet description of spacetime. If one creates a state one cannot simply turn it off as is the case in classical computation, where the effects can be eliminated by simply setting a variable to its natural zero state. The same is valid for spacetime worldsheets. Their history is preserved no matter what we decide to do at some point, and hence one cannot simply "undo" the effects encoded in its history by turning a spacetime worldsheet "off". Ancilla qubits can be used to preserve the global information over entangled states, while performing operations on the original states themselves, using local operations. Without such ancilla qubits, such operations would not be allowed by means of local computation. In an analogous sense let us consider what we are performing in the construction of T-dual theories. Considering the process of introducing the auxiliary gauge fields as analogues to ancillary qubits, we introduce them to encode the gauge symmetry while enforcing a special global structure, namely that the overall curvature vanishes. We prepare and preserve this by means of another auxiliary variable (analogue to a second ancilla qubit). However, while we keep the curvature term zero, and we expect the fields to be pure gauge, there exists global information that has been encoded and shared between the two dual partners. The ancilla qubits preserve it in a sense analogous to entanglement, just applied at a higher level, between two dual theories. The process of "measurement" is basically the result of integrating out the respective fields hence the process of "undoing" the local structures created before via the ancillas. While we undo the local actions, we preserve the "entanglement" which is recovered globally as two dual theories. 
Hence T-duality, appears to be some higher form of entanglement which can be interpreted by means of quantum circuits in some higher category where a functorial relation between the quantum circuits and gauge theories is manifest. 
\section{conclusion}
Quantum information theory has universal validity. By this we usually mean that we can create entangled states of objects and perform universal computations on them. However, there exists a more general notion of universality, one that is related to a categorial and functorial interpretation of quantum information that remains valid not only at the level of entangled photons or electrons, but also at a more abstract and hence more fundamental level, for example in the case of categories that have similar properties as quantum mechanics and hence are, form an axiomatic point of view, already "quantum". An axiomatisation of quantum mechanics has not been performed, but is deeply desired. While not having such a formulation, I assumed that non-separability of spaces of states in simple cartesian products of states is a fundamental aspect of quantum mechanics, and it is valid at the level of the fibre bundles which are also known to encode how a cartesian product fails to be cartesian due to global information. In reality, entangled states are exactly what remains if one eliminates from a quantum space of states all the states that are separable. Hence, this property seems to be rather general. By using this admittedly heuristic approach, one can implement quantum information operations on structures that are more general than our usual quantum states. In fact it is not known if there are other physical systems except the quantum ones that exhibit such non-separability properties. Hence, at least at the level of conjecture, one can assume this to be a fundamental aspect of a system being quantum. On the other side, most likely more axioms will be needed to properly define quantum mechanics. I expect however that this structure defined here will be maintained even when such axioms are added in a consistent way. However, it is also possible that by giving up on some or all of the other axioms, new more general versions of "quantum mechanics" will emerge. I do not know whether those will be relevant in any physical sense. Here, I used the analogue of ancilla qubits to explain a well known procedure of introducing auxiliary fields in a gauge theory and integrating them away, in the same way in which we use ancillas in quantum mechanics to introduce operations we wish to annihilate after their use. Of course, those operations will leave their global, locally undetectable, traces on the combined system, which are here interpreted as a T-duality. 
There is much work left. First I did not discuss the non-abelian case of the manifold isometry, which creates its own complications. Then, it is strongly desirable to refine the concepts and to give a proper definition to the respective operations done, as well as a stronger connection between the gauge theory tricks involved and ancilla qubit constructions in quantum information. At the same time it would be desirable to show some universal properties that would encode more general principles of dualities between various $\sigma$-models and to properly encapsulate them in a categorial approach. If anything, I hope this article will spark new research in the connection between gauge and string theory dualities on one side and quantum information and its categorial interpretations on another side. 
\section{Data Availability Statement}
No Data associated in the manuscript

\end{document}